\documentclass[letterpaper,twocolumn,10pt]{article}
\usepackage{usenix,epsfig,endnotes}
\begin{document}

\date{}

\title{\Large \bf AllReduce Scheduling with Hierarchical Deep Reinforcement Learning}

\author{
  {\rm Yufan Wei} \\
  \texttt{evan.wway@gmail.com}
  \and
  {\rm Michel Liu} \\
  \texttt{mickelliu7@gmail.com}
  \and
  {\rm Wenfei Wu} \\
  \texttt{wenfeiwu@pku.edu.cn}
}

\maketitle

\thispagestyle{empty}

\subsection*{Abstract}
AllReduce is a technique in distributed computing which saw use in many critical applications of deep learning.
Existing methods of AllReduce scheduling oftentimes lack flexibility due to being topology-specific or relying on extensive handcrafted designs that require domain-specific knowledge. 
In this work, we aim to alleviate this inflexibility by proposing a deep-reinforcement-learning(DRL)-based pipeline that can generate AllReduce scheduling for various network topologies without topology-specific design features.
The flow scheduling module of this pipeline consists of two hierarchically-structured DRL policies that work cooperatively to find optimal scheduling.
We showcase the performance of our method compared to the baseline methods on three topologies: \textit{BCube}, \textit{DCell} and \textit{Jellyfish}.
Finally, we contributed a Python-based simulation environment simulating AllReduced scheduling on these network topologies.

\vspace{-1.25em}
\section{Introduction}
\vspace{-0.75em}

Distributed machine learning (DML) is a powerful technique for training large-scale machine learning models by distributing the computations across multiple machines or devices~\cite{NIPS2014_1ff1de77, wan2020rat}. The AllReduce operation is a key component of DML, which is used to aggregate gradients from different machines before updating model parameters. However, the performance of AllReduce operation relies heavily on the underlying network topology and flow scheduling~\cite{wan2020rat}. Existing methods of AllReduce scheduling oftentimes lack flexibility due to being topology-specific or relying on extensive handcrafted designs that require domain-specific knowledge~\cite{bml-2018}.

In this paper, we aim to alleviate this inflexibility by proposing a deep-reinforcement-learning (DRL)-based pipeline that can generate AllReduced scheduling for various network topologies without topology-specific design features. Our pipeline is built upon the framework of Partially Observable Markov Decision Process (POMDP) which is a powerful tool to model decision-making under uncertainty. By formulating the AllReduce scheduling problem as a POMDP, we can use RL algorithms to learn an optimal policy to schedule the workloads.

\begin{figure}[t]
\centering
\includegraphics[width=0.5\textwidth]{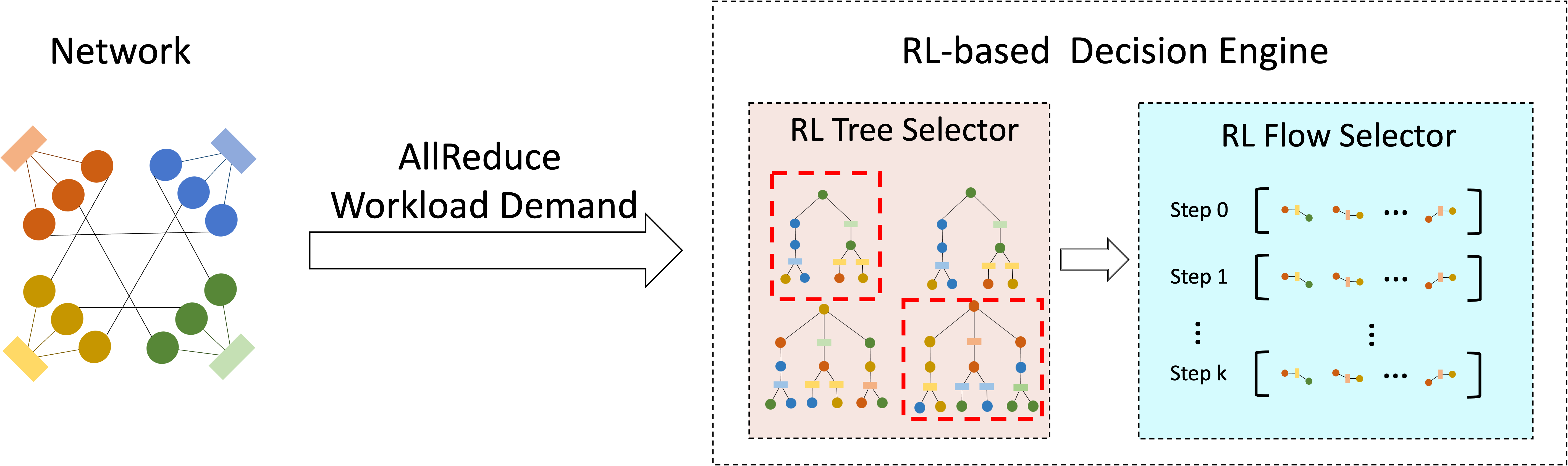} 
\caption{An overview of our method. The objective is to perform AllReduce operation efficiently with the aid of RL-based decision-making models.} 
\label{Fig.main2}
\vspace{-1.5em}
\end{figure}

Our pipeline consists of two hierarchically-structured DRL policies that work cooperatively to find optimal AllReduce scheduling. The higher-level policy determines a set of flow trees and passes to the lower-level policy downstream that outputs a flow scheduling in a sequential manner. By splitting the problem into two separate POMDPs and using a hierarchical structure, we are able to handle the high complexity of the problem and improve the efficiency of the training process.

To evaluate the performance of our proposed method, we have designed a flow-level simulator that consists of two main parts: the topology generator and the flow simulator. The topology generator is capable of generating different network topologies including BCube, DCell, and Jellyfish. The flow simulator is responsible for simulating the flow of data during an AllReduce operation. It uses a flow-tree representation to describe the workloads or flow demand of a server node. The flow simulator also implements the concept of prefix relationships between flows, which allows us to further optimize communication by reducing the amount of data that needs to be transmitted.
We have performed experiments on three network topologies (\textit{BCube}, \textit{DCell} and \textit{Jellyfish}) at various network scales.

Our work is a step forward in the direction of using RL to solve AllReduce scheduling problem in DML, and the pipeline can be easily adapted to other types of network and data flow scheduling. This work could serve as a foundation for future research in this area, enabling more sophisticated and efficient scheduling of AllReduce operations in DML systems.

\vspace{-1.5em}
\section{Related work}\label{related}
\vspace{-0.75em}
\begin{figure*}[!t]
    \centering
    \includegraphics[width=0.8\linewidth]{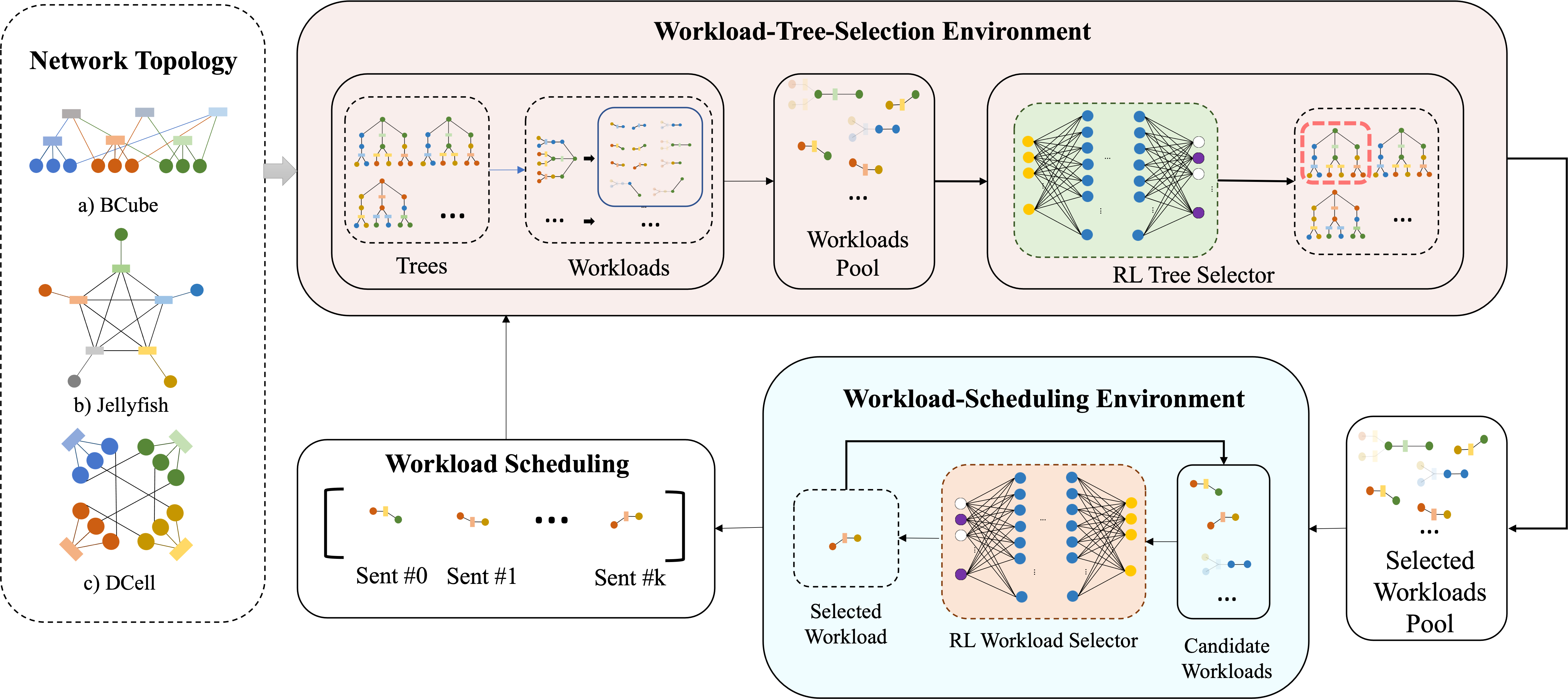}
    \caption{A simplified illustration of our proposed pipeline. We use the flow-tree representation to describe the workloads or flow demand of a server node. The upper-level model determines a set of flow-tree to interact with at a given round, which the set defines a pool of candidates for the lower-level model to perform flow scheduling. The lower-level model determines a valid flow scheduling and sends out the flows to proceed environment to the next round.}
    \vspace{-1em}
    \label{fig:my_label}
\end{figure*}

\paragraph{Date Center Network Topology.} A data center network is a critical component of a data center as it enables the communication between different devices and systems within the data center.
The data center network topology refers to the layout of the various components of the network, including the devices and connections between them.

DCell~\cite{dcell-2008} consists of a hierarchical structure of cells, with each cell containing multiple servers connected by commodity network switches. It is known for its high scalability but can have issues with cross-section bandwidth and latency.
Bcube~\cite{bcube-2009}  utilizes a cube-shaped topology of switches and servers. It provides high bandwidth and low latency but can have higher complexity and cost compared to some other DCN architectures.
Jellyfish~\cite{jellyfish-2012} uses a mesh of switches and links to support efficient communication and high scalability. It provides high bandwidth and low latency but can have higher complexity and cost compared to some other DCN architectures.

\vspace{-1.5em}
\paragraph{AllReduce Operation in DCN.} AllReduce is a distributed computing operation that combines the values of multiple variables from different machines into a single value. It is commonly used in distributed systems to perform collective communication and data aggregation~\cite{patarasuk2009bandwidth, moritz2018ray, lee2020flexreduce, bao2020preemptive }. In the context of machine learning, AllReduce is often used to aggregate gradients from different machines before updating model parameters. There are several different ways to implement the AllReduce operation, including using MPI (Message Passing Interface) or gRPC (Google Remote Procedure Call). Pytorch~\cite{pytorch} and TensorFLow~\cite{tensorflow2015-whitepaper} also includes support for the AllReduce operation.

\paragraph{Applications of Deep Learning in Computer Network.} Many recent works have successfully applied deep learning (DL) techniques to solve research problems in the field of computer networks. \textit{Chen et al.}~\cite{Chen2018-ls} trained reinforcement learning policies to schedule short-long flows for traffic optimization in large networks. \textit{Zhu et al.}~\cite{Zhu2021-ra} used Deep-RL to optimize the search space of ILP algorithm for network planning. \textit{xWeaver}~\cite{Wang2018-av} uses deep convolution networks to generate optimized topology configuration specific to a target data center.

\vspace{-1.3em}
\section{Preliminaries}
\vspace{-0.5em}
\paragraph{Distributed Machine Learning} 
Distributed Machine Learning (DML) is a technique used to train large machine learning models by distributing the computations across multiple machines or devices. There are two main types of DML: data-parallel and model-parallel. 
In this work, we focus on the data parallel scenario, where the whole data is split into smaller chunks, and each machine or device is responsible for training on its own chunk of the data. The gradients are then averaged across all machines or devices, and the model is updated accordingly.


\begin{table}[]
\centering
\resizebox{\columnwidth}{!}{%
\begin{tabular}{|c|l|}
\hline 
\textbf{Notation}&\textbf{Definition}\\
\hline 
$N$&The total number of servers in topology\\
\hline 
$P$&The total number of workloads\\
\hline 
$k$&The number of gradient pieces\\
\hline  
$T_F$&The theoretical time to transmit full workloads\\
\hline  
$T_S$&The theoretical time to transmit a single workload\\
\hline  
$N_{\mathrm{on}}$&The number of workloads on stream\\
\hline  
$N_{\mathrm{phy}}$&The number of physical links\\
\hline 
\end{tabular}
}
\vspace{-1em}
\caption{Notations used throughout the paper}
\vspace{-1.5em}
\label{tab:notations}
\end{table}


In this paper, we summarize the various notation used throughout the text in Table~\ref{tab:notations}. The notation $N$ denotes the number of server machines in a DML network, while $N_{\mathrm{phy}}$ denotes the number of physical links within the network. The symbol $P$ is used to represent the size of the entire gradients for the model. Additionally, we use $T_F$ to represent the theoretical time required to transmit the entire gradients at full speed. In order to optimize the utilization of the link capacity of the synchronization process, many algorithms choose to divide the gradients into $k$ pieces, each of which has a cost of $\frac{T_F}{k}$, represented by $T_S$. Given limitations in physical link capacity, only a limited number of gradient pieces, denoted by $N_{\mathrm{phy}}$, can be transmitted concurrently as workloads during the synchronization process, resulting in the potential for multiple \textit{epochs} within a single synchronization process. It is natural to consider the ratio of $N_{\mathrm{on}}$ to $N_{\mathrm{phy}}$ as a criterion in evaluating the performance of the algorithms.


\vspace{-1.25em}
\paragraph{AllReduce operation.} 
AllReduce is a collective communication operation that is commonly used in DML for parameter and gradients synchronization. It allows multiple machines to collectively perform a reduction operation on their local data, and then broadcast the result back to all machines. For example, in a N-server DML system, each server has its own gradient after local computing, whose size is $\frac{P}{N}$ as mentioned paragraph. The AllReduce operation is executed in two stages, known as reduce-scatter and all-gather.

\vspace{-1.25em}
\paragraph{Partially Observable Markov Decision Problem (POMDP).}
Markov Decision Process (MDP) is a mathematical framework for modeling decision-making that forms the basis of any reinforcement learning problem. POMDP is a variant of MDP where the agent only has partial observability over the true states of the environment. $\mathcal{S}$ describes the state space of the environment, $\mathcal{O}$ is the agent's observation space and $\mathcal{A}$ is the set of all possible actions of the agent. At each timestep $t$, the current state of the environment is $s_t\in\mathcal{S}$ and the agent produces an observation $o_t\in\mathcal{O}$ which depends on the current state. To choose actions, the agent uses a stochastic policy $\pi_\theta$ parameterized by $\theta$ and then samples the current action from this policy, $a_t\sim\pi_\theta(o_t)$. Action $a_t$ causes the environment to transition to the next state $s_{t+1}$ with probability $\mathcal{T}(s_{t+1}|s_t,a_t)$, where $\mathcal{T}$ is a state transition function. The agent will receive a reward $r(s_t,a_t)$ as a function of the current state-action pair. The goal of the agent is to maximize the expected return within time horizon $T$, $R_T=\mathop{\mathbb{E}}[\sum_{t=0}^T \gamma^t r_t]$, which $\gamma$ is the discount factor.


\vspace{-1.5em}
\section{Method}
\vspace{-0.75em}
In the following section, we will first introduce our AllReduce Flow Simulation environment and network topology generator used in this study. Next, we will describe our formulations of breaking down the AllReduce scheduling problem into two separate POMDPs, and our iterative training scheme to iteratively train two RL agents based on these two POMDPs. 

\vspace{-1.2em}
\subsection{AllReduce Flow Simulation and Topology Generation}
\vspace{-0.6em}

\paragraph{Simulator Design.}
In order to evaluate the performance of our proposed method and baselines, we have designed a flow level simulator that consists of two main parts: the topology generator and the flow simulator. 

The topology generator is capable of generating different network topologies including BCube, DCell and Jellyfish. The flow simulator, which is the novel aspect of our proposed method, is responsible for simulating the flow of data during an AllReduce operation.

\vspace{-1.5em}
\paragraph{Workload Tree Model.}
Our flow simulator utilizes the Parameter Server architecture for distributed systems, where a centralized server manages the system's parameters and coordinates the flow of gradients between the nodes in the system. Unlike traditional methods that rely on logical trees to represent the flow of data, our simulator employs a node tree to visualize the physical-link level transmission dynamics. The node tree is constructed by representing the server nodes as individual nodes, and the physical routes as edges, allowing for an in-depth analysis of the communication patterns at the physical layer. Additionally, the simulator generates a workload tree, which represents all workloads generated from the node tree with one specific node as the root node. A workload tree organizes all workloads with the same data destination. By using the tree model, the flow simulator is able to accurately model the communication patterns during an AllReduce operation, enabling a more detailed and accurate evaluation of any proposed methods.

\vspace{-1.5em}
\paragraph{Parameter server and Merge operation.}

To reduce the number of workloads and improve the performance of the network, we propose a merge operation on the branches of the workload tree. As previously mentioned, the root node performs computation tasks such as summarization after receiving all prefix workloads. By merging branches that are pointing to the same server node and have the same data destination, we are able to group multiple workloads together, thus reducing the number of transmissions required to complete the AllReduce operation. This also leads to a compacting of the workload tree, reducing the transmission pressure for the synchronization process. It is important to note that branches that end in switch nodes cannot be merged, as switches are not capable of performing aggregation. By reducing the number of branches through the merge operation, the AllReduce operation can be completed more efficiently, thus improving the performance of the network.
\begin{figure}[tb]
    \centering
    \includegraphics[width=0.45\textwidth]{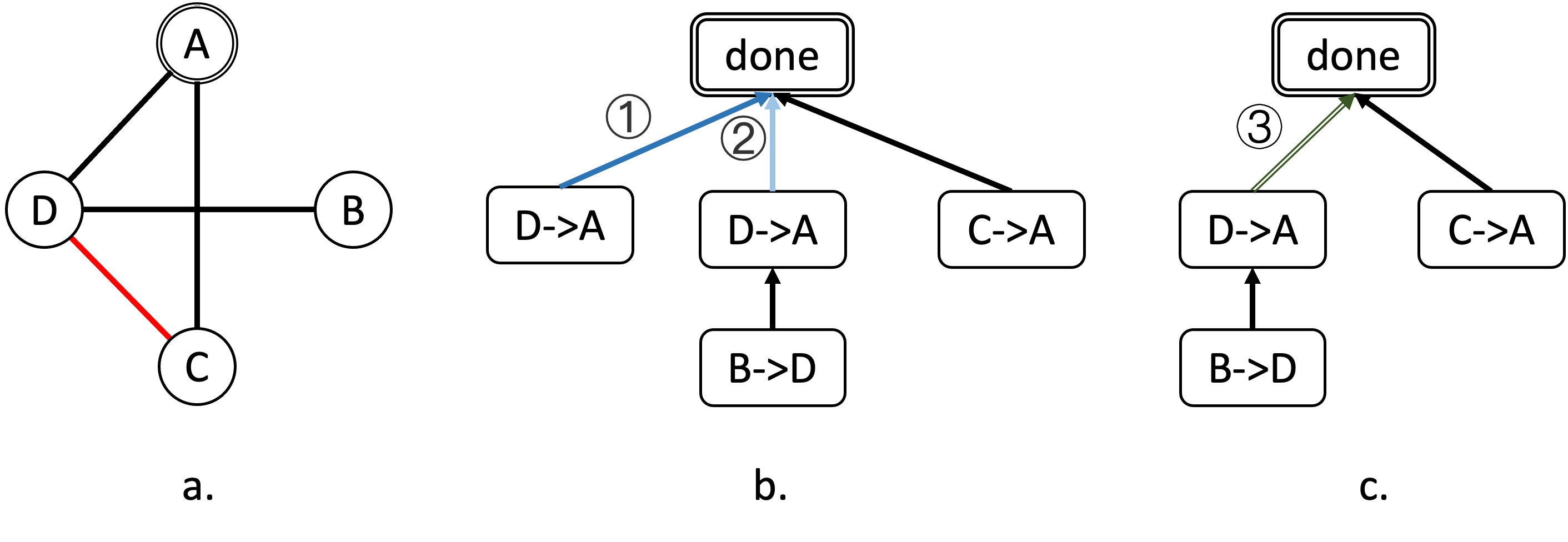}
    \vspace{-1.5em}
    \caption{Example for Parameter server and merge operation}
    \label{fig:node3}
    \vspace{-1.5em}
\end{figure}

As an example, Figure ~\ref{fig:node3} illustrates a graph generated from a topology, with hosts A, B, C, and D in the physical link. Through the application of the shortest-path routing algorithm, it can be observed that the edge C-D can be deleted as there are no flows between them. Subgraph B depicts the workload tree of the example topology, and it can be observed that workloads \textcircled{1} and \textcircled{2} share the same physical link. As a result, they are merged as shown in subgraph c, resulting in a tree with fewer edges.

\vspace{-1.25em}
\paragraph{Workload.}
The flow simulator generates a specific \textit{workload structure} which includes information about the root node, indicating the tree it belongs to, source and destination, specifying the physical link it intends to occupy, and prefix relationships between workloads. A \textit{workload} is only available for transmission when the following conditions are met: either their prefix flows have been fully transmitted or if no such flows exist, or it does not conflict with other impending workloads. The objective for the RL agent is to complete all workloads with a minimum number of rounds.

\vspace{-1.25em}
\subsection{Formulation}\label{sec:formulation}
\vspace{-.5em}

\paragraph{Issues with Conventional Formulation.}
The conventional formulation of a single MDP commands the RL agent to output an action in the form of a valid flow scheduling in one-shot for each round. 
Naturally, the action space of this model will scale exponentially with the number of total flows of a given topology. This leads to a notoriously hard exploration problem due to the large space of exploration. Since there are so many possible pairings between states and actions, it takes significant amounts of trial and error to assign credit between successful states and actions. The hard exploration problem results in difficulties in credit assignment, which is very likely leading to failure in training the RL agent to perform tasks. 

\vspace{-1.25em}
\paragraph{Two Hierarchical POMDPs.} 
To alleviate the hard exploration problem and difficulties in credit assignment, we propose to divide the AllReduce scheduling into two sub-problems with a component that focuses on long-term planning and another component that focuses on producing valid scheduling. 

The first problem is \textbf{Flow-Tree Selection} which aims to narrow down the choices for flow routing at a given round. Formally defined as a POMDP, the agent makes an observation of the current state of the network, which includes information such as routed flows, flows available for routing, network pressure, last performed action by the agent, \etc. The agent's action $a$ defines as a multi-hot vector of length equals the total number of flow-trees. The selected set of flow-trees is a subset of all flow-trees of a given topology, which manages the set of candidates needed to be considered by the downstream workload-scheduling agent.



The second problem is \textbf{Workload Scheduling}. The agent is tasked to determine the set of non-conflicting workloads for each round. This problem has a different notion of \textit{time} compared to the first problem. Whereas the first problem defines \textit{time} in terms of ``number of rounds'', this problem uses the within-round steps as the notion of \textit{time} and therefore determines a valid workload scheduling through a sequential decision process. The agent then sequentially outputs the selected workload for every step until a termination condition is met (\ie~no more non-conflicting workload is available to be selected), and by combining the outputs form a valid workload scheduling. For example, suppose in a given round the workload-scheduling agent ran for $M$ step. At each step the agent outputs an action $a^m_\text{WS} \in \mathcal{A}$, $m=1,2,\dots,M$, and the resulting flow-scheduling is a sequence $[a^0_\text{WS}, a^1_\text{WS},\dots, a^M_\text{WS} ]$.

\vspace{-1.25em}
\paragraph{Hierarchical RL (HRL).} Problem-solving with HRL is a long-standing topic of study in the field of reinforcement learning~\cite{sutton1999between, pmlr-v70-vezhnevets17a, bacon2017option}. It is a divide-and-conquer approach to decompose a complex problem into simpler sub-problems and solve them accordingly. AllReduce scheduling is produced by an RL decision engine consisting of two hierarchical RL agents. The upper-level component is the agent trained with the \textbf{Flow-Tree Selection} task and the lower-level component is the agent trained with the \textbf{Workload Scheduling} task. The upper-level \textit{``manager''} agent prioritizes long-term planning, as the agent needs to consider the number of shared paths between flow-trees to ensure less congestion during flow sending done at the lower level. At each round, the \textit{``manager''} agent selects a pool of workloads from all remaining workloads, which reduces the search space for the lower-level agent and potentially leads to more optimal scheduling. Next, the lower-level \textit{``worker''} agent prioritizes maximum flow. Given a selected pool of available workloads, the \textit{``worker''} agent sequentially decides the flow to send at each step of the round, and the resulting sequence of sent flows is a flow scheduling for the round.  

\begin{table*}[h]
\centering
\resizebox{0.8\textwidth}{!}{%
\begin{tabular}{|l|ccc|ccc|ccc|}
\hline
\multicolumn{1}{|c|}{} &
  \multicolumn{3}{c|}{\textbf{BCube}} &
  \multicolumn{3}{c|}{\textbf{DCell}} &
  \multicolumn{3}{c|}{\textbf{Jellyfish}} \\ \hline
\rowcolor[HTML]{E8E8E8} 
\cellcolor[HTML]{E8E8E8}\textit{\begin{tabular}[c]{@{}l@{}} $(N_\text{Node}, N_\text{Edge})$\end{tabular}} &
  \multicolumn{1}{c|}{\cellcolor[HTML]{E8E8E8}\textit{(15, 18)}} &
  \multicolumn{1}{c|}{\cellcolor[HTML]{E8E8E8}\textit{(24, 32)}} &
  \cellcolor[HTML]{E8E8E8}\textit{(35, 50)} &
  \multicolumn{1}{c|}{\cellcolor[HTML]{E8E8E8}\textit{(25, 30)}} &
  \multicolumn{1}{c|}{\cellcolor[HTML]{E8E8E8}\textit{(36, 45)}} &
  \textit{(49, 63)} &
  \multicolumn{1}{c|}{\cellcolor[HTML]{E8E8E8}\textit{(20, 30)}} &
  \multicolumn{1}{c|}{\cellcolor[HTML]{E8E8E8}\textit{(30, 45)}} &
  \textit{(40, 59)} \\
\rowcolor[HTML]{E8E8E8} 
\textit{Number of Workloads} &
  \multicolumn{1}{c|}{\cellcolor[HTML]{E8E8E8}\textit{144}} &
  \multicolumn{1}{c|}{\cellcolor[HTML]{E8E8E8}\textit{240}} &
  \textit{1200} &
  \multicolumn{1}{c|}{\cellcolor[HTML]{E8E8E8}\textit{380}} &
  \multicolumn{1}{c|}{\cellcolor[HTML]{E8E8E8}\textit{870}} &
  \textit{1722} &
  \multicolumn{1}{c|}{\cellcolor[HTML]{E8E8E8}\textit{180}} &
  \multicolumn{1}{c|}{\cellcolor[HTML]{E8E8E8}\textit{420}} &
  \textit{760} \\ \hline
Parameter Server (PS) &
  \multicolumn{1}{c|}{16.8} &
  \multicolumn{1}{c|}{31.8} &
  51.6 &
  \multicolumn{1}{c|}{30.0} &
  \multicolumn{1}{c|}{48.4} &
  71.2 &
  \multicolumn{1}{c|}{23.0} &
  \multicolumn{1}{c|}{\textbf{36.0}} &
  \textbf{51.2} \\
Ring AllReduce &
  \multicolumn{1}{c|}{18.0} &
  \multicolumn{1}{c|}{64.0} &
  150.0 &
  \multicolumn{1}{c|}{47.1} &
  \multicolumn{1}{c|}{75.9} &
  112.3 &
  \multicolumn{1}{c|}{40.0} &
  \multicolumn{1}{c|}{69.6} &
  80.0 \\
\textbf{Ours (RL-Based)} &
  \multicolumn{1}{c|}{\textbf{10.2}} &
  \multicolumn{1}{c|}{\textbf{20.8}} &
  \textbf{34.7} &
  \multicolumn{1}{c|}{\textbf{23.2}} &
  \multicolumn{1}{c|}{\textbf{33.8}} &
  \textbf{48.0} &
  \multicolumn{1}{c|}{\textbf{22.7}} &
  \multicolumn{1}{c|}{39.9} &
  62.2 \\ \hline
\end{tabular}%
}
\caption{A performance comparison between three methods on three topologies. \textbf{Unit of Measurement:} number of rounds}
\vspace{-0.5em}
\label{tab:exp-table}
\end{table*}

\vspace{-1.2em}
\subsection{Iterative Training Scheme} 
\vspace{-0.5em}
It is not difficult to observe that these POMDPs have a hierarchical relationship, in which the state transition in the \textbf{Flow-Tree Selection} environment will depend on the behaviours of the agent in the \textbf{Workload Scheduling} environment. However, it is technically difficult to train two agents in these two environments simultaneously, since two POMDPs operate on different notions of \textit{time} and these agents have different action spaces. Recall that \textbf{Flow-Tree Selection} operates on workload-sending rounds and outputs a set of selected flow-tree at each round (\ie~$a^t_\text{FTS}$), while \textbf{Worklord Scheduling} operates on within-round steps for determining which individual workload to send at each step (\ie~$a^m_\text{WS}$). Hence, the trajectories collected in these environments cannot be used interchangeably to co-train the other agent from the different environments. Concretely, 
\vspace{-.75em}
\begin{equation}\label{eqn:tau-tree}
    \tau_\text{FTS}=\left[ (o_\text{FTS}^t, o_\text{FTS}^{t+1}, a^t_\text{FTS}, r_\text{FTS}^t) \right]_{t=1}^T,
    \, \forall \tau_\text{FTS}\in \mathcal{D}_\text{FTS} 
\end{equation}
\vspace{-1em}
\begin{equation}\label{eqn:tau-flow}
    \tau_\text{WS}=\left[ (o_\text{WS}^m, o_\text{WS}^{m+1}, a^m_\text{WS}, r_\text{WS}^m) \right]_{m=1}^M,
    \, \forall \tau_\text{WS}\in \mathcal{D}_\text{WS} 
\end{equation}
\vspace{-.75em}

\algrenewcommand\algorithmicindent{1em}%
\begin{algorithm}[h]
  \caption{Hierarchical-DRL Training Scheme for Learning Optimal AllReduce Scheduling}
  \small
  \begin{algorithmic}[1]
    \State\textbf{Initialize:} flow-tree selection policy $\pi_\text{tree}$, workload-scheduling policy $\pi_\text{flow}$, training iterators $I, J, K$
    \State $\pi_\text{tree}, \pi_\text{flow} \gets$ random initialization 
    \For{$i=1,2,\dots,I$}
        \For{$j=1,\dots,J$}\Comment{\textbf{Flow-Tree Selection (FTS)}} 
            \State Freeze the parameters of $\pi_\text{flow}$
            \State Collect dataset $\mathcal{D}_\text{FTS}$ in using ($\pi_\text{tree}$, $\pi_\text{flow}$), for each round:
            \begin{itemize}
            \setlength{\itemindent}{.05in}
                \item $\pi_\text{flow}$ outputs a flow-tree selection $a^t_\text{FTS}$ given the set of remaining workloads $W_t$ and the set of all flow-trees $F$
                \item $a^t_\text{FTS}$ selects a pool of to-be-sent workloads $w_t$, $w_t\subseteq W_t$
                \item $\pi_\text{tree}$ outputs $[a^0_\text{WS}, \dots, a^M_\text{WS}  ]$, a valid scheduling of $w_t$
                \item Obtain next state $s_{t+1}$, update remaining workloads $w_{t+1}$  
            \end{itemize}
            \State Train $\pi_\text{tree}$ on $\mathcal{D}_\text{FTS}$~(\ref{eqn:tau-tree})
        \EndFor
        \For{$k=1,\dots,K$}\Comment{\textbf{Workload Scheduling (WS)}} 
            \State Freeze the parameters of $\pi_\text{tree}$
            \State Collect dataset $\mathcal{D}_\text{WS}$ using $(\pi_\text{tree}$, $\pi_\text{flow})$, similar to step (6)
            \State Train $\pi_\text{tree}$ on $\mathcal{D}_\text{WS}$~(\ref{eqn:tau-flow})
        \EndFor
    \EndFor
  \end{algorithmic}
\end{algorithm}

\vspace{-1.5em}
\subsection{Reward}
\vspace{-0.5em}
We proposed separate reward functions for both POMDPs. This is due to the differences in the nature of the tasks, particularly their differences in optimization objectives.

\vspace{-1.5em}
\paragraph{Training Flow-Tree Selection.} 
As this is the POMDP served as the environment of our upper-level agent, where the majority of the decisions associated with macroscopic planning of flow routing occur, we wish to encourage the agent to propose a sequence of flow-tree selection such that the overall number of rounds is minimized. The reward function $R^{\text{FTS}}=R_{\mathrm{Dense}}^{\text{FTS}}+R_{\mathrm{Stage}}^{\text{FTS}}$ consists of a dense component that provides feedback at every timestep and a stage reward component that provides feedback upon the agent reaching a specific stage.
\vspace{-0.75em}
\begin{equation}\label{eqn:reward_fts_dense}
    R_{\mathrm{Dense}}^{\text{FTS}} = \frac{\text{Sent Flows}}{\text{Total Flows}} + 0.1\times\frac{\text{Flow-Trees Selected}}{\text{Total Flow-Trees}}
\end{equation}
\vspace{-1.2em}
\begin{equation}\label{eqn:reward_fts_stage}
    R_{\mathrm{Stage}}^{\text{FTS}} = 
        \begin{cases}
        10 & \text{if } \textbf{Done}\\
        -\frac{\text{Total Flow-Trees}}{\text{Total Flows}} & \text{otherwise }
    \end{cases}
\end{equation}
\vspace{-0.75em}

$R_{\mathrm{Dense}}^{\text{FTS}}$ in Eqn.~\ref{eqn:reward_fts_dense} encourages the agent to output sets of flow-tree selections that maximize the throughput from the lower-level agent measured by the number of sent flow per round. 

\vspace{-1.25em}
\paragraph{Training Flow Scheduling.} The lower-level agent is trained in the POMDP associated with flow scheduling given a set of flow-tree selected by the upper-level agent. The lower-level agent is tasked to generate a valid flow scheduling. Naturally, the reward function should encourage the agent to schedule as many workloads as possible at each round. 
\vspace{-0.5em}
\begin{equation}\label{eqn:reward_fs}
    R^{\text{FS}} = \frac{1}{\text{Total Flows}}
\end{equation}

\vspace{-1.5em}
\section{Experiment}
\vspace{-.5em}

We pick the average number of rounds to complete all workloads as the evaluation metric. We introduce two other AllReduce scheduling methods to be used as the baselines in our experiments:

\vspace{-1.25em}
\paragraph{Parameter Server.} The Parameter Server~\cite{44634} method is implemented in a P2P fashion, which means each server is not only a parameter server but also a centralized server. The worker nodes send their gradients to the parameter server, which performs the computation on the gradients and sends the updated parameters back to the worker nodes.

\vspace{-1.25em}
\paragraph{Ring AllReduce.}
Ring AllReduce~\cite{gibiansky2017bringing} is another widely used method for flow scheduling in distributed systems, particularly in high-performance computing environments. This method is based on the idea of organizing the nodes in a network into a logical ring topology, where each node is connected to two other nodes, and the first node is connected to the last node. In the reduce-scatter stage of the AllReduce operation, each node sends its data to one neighbour and receives data from the other neighbour along the ring.

\vspace{-1.25em}
\paragraph{Analysis.}
Table~\ref{tab:exp-table} showcases that  our method consistently outperforms PS and Ring AllReduce in \textit{Bcube} and \textit{DCell} by a significant margin. However, PS outperforms our method in two out of three jellyfish topologies. According to this observation, we conclude that our RL workflow works better for server-centric topologies such as \textit{BCube} and \textit{DCell}. In these topologies, switches and servers are interconnected with more links therefore many possible flow paths share a common data destination that could be optimized by the merge operation. Whereas the jellyflow network is a specialty network where at the center is a cluster of interconnected switches and the server nodes are scattered at the edge. This means the majority of the connections within a jellyfish network are switch-to-switch connections. As we assume that switch nodes have no computational capabilities, this leaves little room for optimization by the merge operation.

\vspace{-1.25em}
\section{Conclusion}
\vspace{-0.75em}
In this paper, we proposed a DRL-based pipeline for AllReduce scheduling in various network topologies. We proposed the merge operation to reduce repeating transmission of flow data. We divide the AllReduce scheduling task into two simpler sub-problems handled by two different RL agents, and a valid flow scheduling can be produced in an end-to-end manner by the collaborative work of these two agents.




\newpage
{\footnotesize \bibliographystyle{acm}
\bibliography{reference}}


\end{document}